\pdfoutput=1

\documentclass[aps,prc,superscriptaddress,showpacs,nofootinbib,floatfix,twocolumn]{revtex4}

\usepackage{graphicx}	

\usepackage{xspace}	
\usepackage{amsmath}

\newcommand{\pt}{\mbox{$p_{T}$}\xspace}
\newcommand{\rg}{\mbox{$R_g$}\xspace}

\newcommand{\rdau}{\mbox{$R_{\rm{dAu}}$}\xspace}

\newcommand{\dau}{\mbox{{\rm d}+{\rm Au}}\xspace}  
\newcommand{\dA}{\mbox{{\rm d}+{\it A}}\xspace}  

\newcommand{\sqs}{\mbox{$\sqrt{s}$}\xspace}
\newcommand{\sqsn}{\mbox{$\sqrt{s_{_{NN}}}$}\xspace}
\newcommand{\jpsi}{\mbox{$J/\psi$}\xspace}  

\newcommand{\ccbar}{\mbox{$c\overline{c}$}\xspace}

\newcommand{\sbr}{\mbox{$\sigma_{\rm abs}$}\xspace}

\newcommand{\rt}{\mbox{$r_{T}$}\xspace}

\newcommand{\qsq}{\mbox{$Q^2$}\xspace}

\newcommand{\chisqm}{\mbox{$\overline{\chi}^{2}$}\xspace}

\begin{document}

\title{Impact parameter dependence of the nuclear modification of \jpsi  production in d+Au collisions at \sqsn  = 200 GeV}

\author{D. McGlinchey}
\author{A. D. Frawley}
\affiliation{Physics Department, Florida State University, Tallahassee, FL 
32306, USA}
\email{dmcglinchey@fsu.edu}
\email{afrawley@fsu.edu}

\author{R. Vogt}
\affiliation{Physics Division, Lawrence Livermore National Laboratory, 
Livermore, CA 94551, USA}
\affiliation{Physics Department, University of California at Davis, Davis, CA 
95616, USA}
\date{\today}

\begin{abstract}
The centrality dependence of $\sqsn$ = 200 GeV \dau \jpsi data, measured in 
12 rapidity bins that span 
$-2.2 < y < 2.4$, has been fitted using a model containing an effective absorption 
cross section combined with EPS09 NLO 
shadowing. The centrality dependence of the shadowing contribution was allowed 
to vary nonlinearly, employing a variety of assumptions, in an effort to 
explore the limits of what can be determined from the data. The impact parameter
dependencies of the effective absorption cross section and the 
shadowing parameterization are sufficiently distinct to be determined 
separately. It is found that the onset of
shadowing is a highly nonlinear function of impact parameter. 
The mid and 
backward rapidity absorption cross sections are compared with lower 
energy data and, for times of 0.05 fm/$c$ or greater, data over 
a broad range of collision energies and rapidities are well described by 
a model in which the absorption cross section depends only on time spent in the nucleus.
\end{abstract}



\maketitle



\section{Introduction}

The modification of the gluon distributions in nuclear targets in high energy 
collisions, referred to here as gluon shadowing, is inherently interesting 
because of what it can teach us about the behavior of gluons at low Bjorken 
momentum fraction, $x$, where the gluon densities are high and
saturation effects are expected to become important~\cite{Gelis:2010nm}. 
In addition, the modification of parton distributions in nuclei determines the 
initial conditions in a high energy nuclear collision.  The initial
conditions must be sufficently well understood before final-state hot matter 
effects can be isolated.

Parameterizations of the dependence of nuclear-modified parton distribution 
functions (nPDFs) on $x$ and squared momentum transfer, $Q^2$, have been 
extracted by several groups from data that 
include deep inelastic electron-nucleus scattering (DIS) and Drell-Yan (DY) 
dilepton production in $p+A$ collisions.  The DIS and DY data together provide 
strong constraints on valence and sea quark 
modifications~\cite{Hirai:2007sx,Kovarik:2010uv,deFlorian:2011fp,Eskola:2009uj}.
Including neutrino-induced DIS 
data from heavy targets can discriminate between quarks and 
antiquarks~\cite{Kovarik:2010uv,deFlorian:2011fp}. Inclusive pion production 
data from RHIC have also been incorporated to better constrain the gluon 
modifications~\cite{Eskola:2009uj,deFlorian:2011fp}. 

The measurements used to extract the nPDFs cited 
above~\cite{Hirai:2007sx,Kovarik:2010uv,deFlorian:2011fp,Eskola:2009uj} were 
all averaged over impact parameter. Therefore
these nPDFs represent the parton modification averaged over the entire nucleus. 
If the modification of these nPDFs is desired as a function of the 
impact parameter, a specific dependence has been assumed~\cite{Klein:2003dj}.
A different approach, employing Gribov theory and incorporating diffractive
data, allows the spatial information to be retained~\cite{Frankfurt:2003zd}.

Recently, the impact parameter dependence of the EPS09~\cite{Eskola:2009uj} and 
EKS98~\cite{Eskola:1998df} nPDFs has been 
parameterized~\cite{Helenius:2012wd} using the target 
mass dependence of the EPS09 and EKS98 parameter sets themselves. 
Terms up to fourth order in the nuclear thickness were necessary to produce
$A$-independent coefficients. 

In this paper, we address the impact parameter dependence of gluon shadowing 
in a different way, using the collision centrality and 
rapidity dependence of the \jpsi yields measured in $\sqsn=200$ GeV d+Au 
collisions at RHIC~\cite{Adare:2010fn}. We were motivated by the 
observation~\cite{Adare:2010fn,Nagle:2010ix} 
that the onset of \jpsi suppression at forward rapidity suggests a quadratic
or higher dependence on the nuclear thickness function at impact parameter \rt, 
$T_A(\rt)$. 

Gluon-gluon interactions dominate \jpsi production in high-energy hadronic 
collisions. Therefore, \jpsi production in $p$(d)$+A$ collisions 
must reflect the gluon modification in the nuclear target.  However, the 
measured modifications of \jpsi yields in $p$(d)$+A$ collisions relative to 
$p+p$ collisions are also sensitive to the breakup of bound $c\overline{c}$ 
pairs by collisions with nucleons as the pairs pass through the medium. This 
effect, as well as effects due to any processes aside from shadowing, are 
usually parameterized by an effective absorption cross section, \sbr, fitted to
the measured data (see {\it e.g.} Ref.~\cite{Lourenco:2008sk}). 
The main goal of this work is to determine whether 
the impact parameter dependence of shadowing could be separated 
from the effects embodied in \sbr. Because the 
magnitude of the effect due to \sbr depends exponentially on 
nuclear thickness for a constant \sbr, such separation may be possible if 
shadowing has a stronger thickness dependence than absorption.

\section{Model Inputs and Fitting Procedure}

We assume that the shadowing modification, integrated over all \rt, could 
be described by the EPS09 NLO 
gluon modification~\cite{Eskola:2009uj} and fit both the 
\rt dependence of shadowing and the magnitude of \sbr to the data.  
We tested two assumed forms for the \rt dependence of shadowing.
Each of the two postulated behaviors had one or two parameters that were 
adjusted to the data, along with the magnitude of \sbr.  A Glauber Monte 
Carlo calculation~\cite{Miller:2007ri} was
employed to compare the modification of the \jpsi yields to the PHENIX 
d+Au data, which are averaged over four centrality bins and twelve 
rapidity bins and then integrated over all \pt. 

The Glauber Monte Carlo calculation allows the modification, 
calculated for individual nucleon-nucleus collisions, to be correctly averaged 
and integrated over collision centrality (related to impact parameter), 
rapidity and \pt.  It also 
accounts for the effects of trigger efficiency in peripheral events. 
The Glauber parameters used here are identical to those used by PHENIX 
when calculating the experimental centrality distributions~\cite{Adare:2010fn}.
For that reason, we found we could drop the uncertainties from
the measured data points that are associated with the mean number of binary 
collisions because they are common to both the data and the calculations. 
The Woods-Saxon nuclear density distribution has a radius of 6.34 
fm and a diffuseness of 0.54 fm.  It is assumed that the nuclear modification 
of the deuteron is negligible. The baseline $J/\psi$ \pt and rapidity 
distributions used in the calculation were the $p+p$ distributions measured by 
PHENIX~\cite{Adare:2012qf}.

\begin{figure}[htb]
 \includegraphics[width=1.0\linewidth]{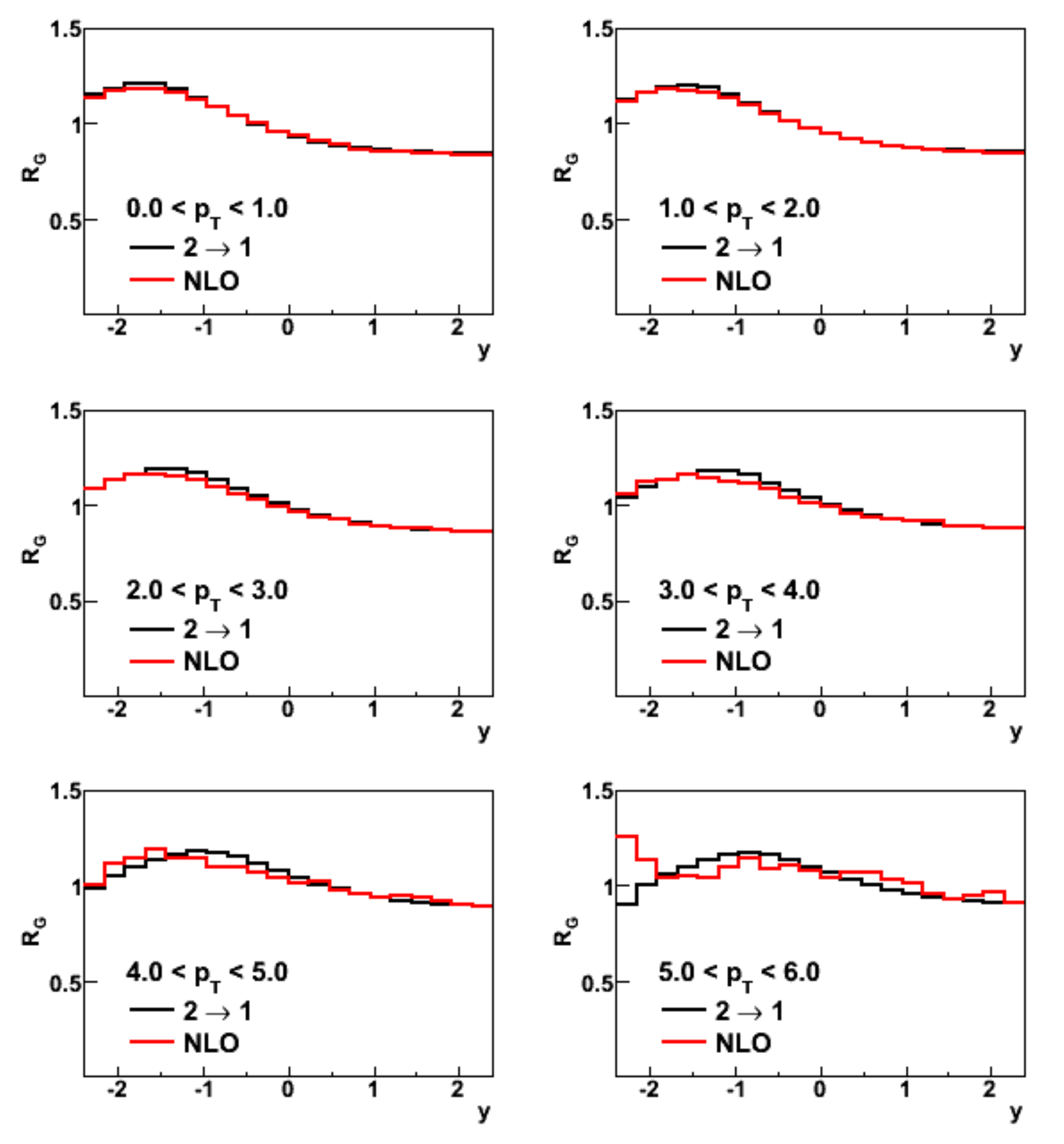}
  \includegraphics[width=1.0\linewidth]{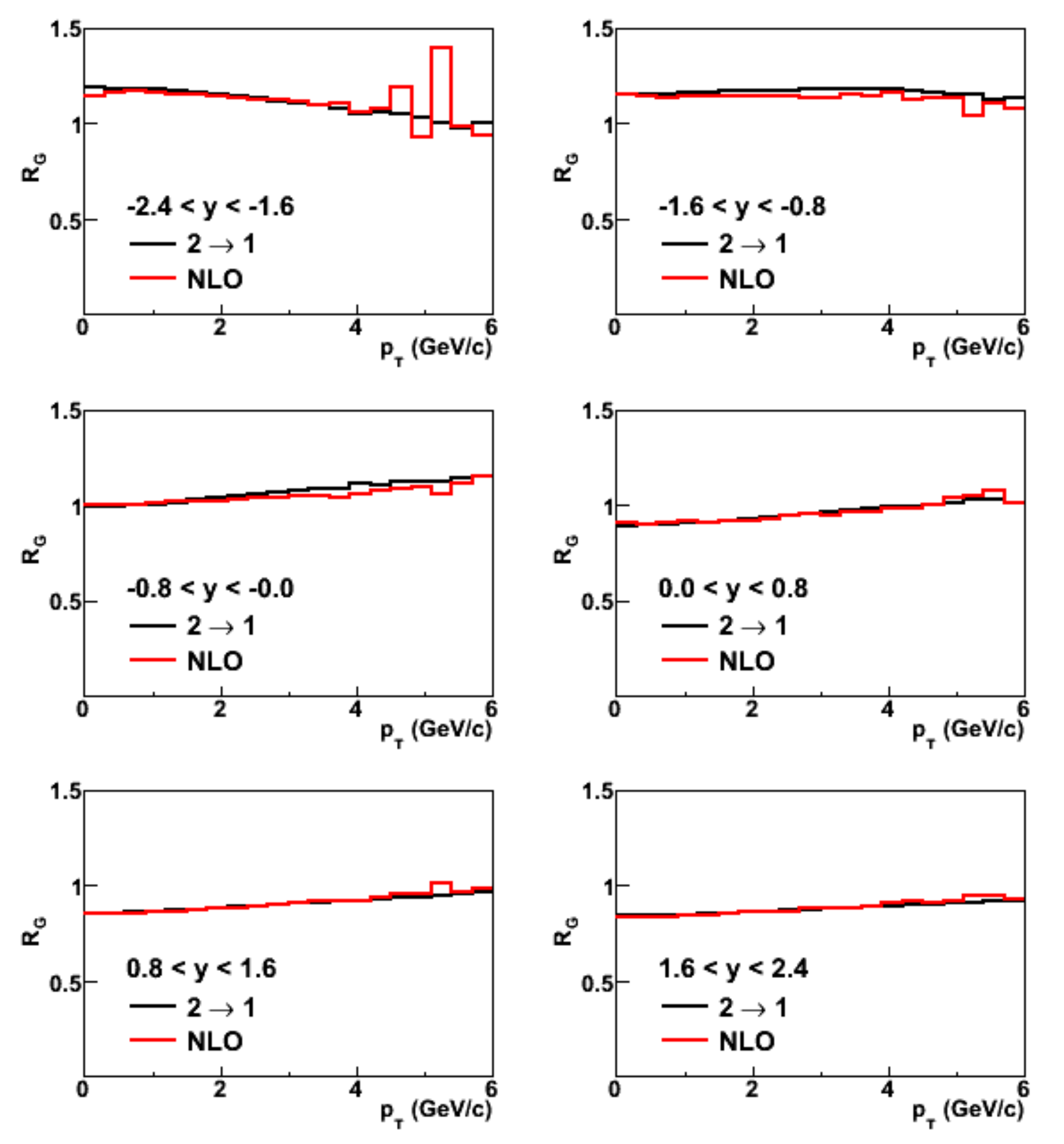}
 \caption{ 
  Comparison of the gluon modification as a function of rapidity and
  $p_T$ obtained from EPS09 NLO using $x$ and $Q^2$ values from Eq.~(\ref{eqn:xq2})
 and from the CEM calculation~\cite{Gavai:1994in}, NLO  
  in the total cross section.
   }
 \label{fig:kinematics}
\end{figure}

The values of the target momentum fraction, $x_2$, and squared momentum 
transfer, $Q^2$, were assumed to obey approximate $2 \rightarrow 1$ 
kinematics as functions of $J/\psi$ rapidity and transverse momentum:
\begin{eqnarray}
x_2 = \frac{\sqrt{{M^2} + p^2_{T}}}{\sqsn } e^{-y} \, \, ,
\,\,\,\,\,  
\qsq = M^2 + p^2_{T}  \, \, ,
\label{eqn:xq2}
\end{eqnarray}
where $M$ is the $J/\psi$ mass.
The $2 \rightarrow 1$ kinematics are not strictly correct since a high $p_T$
$J/\psi$ requires production of an associated hard parton. However, 
Eq.~(\ref{eqn:xq2}) differs from exact $2 \rightarrow 1$
kinematics since the $p_T$ of the $J/\psi$ is finite.  This approximation 
is close to the inclusive $J/\psi$ kinematics in the color evaporation
model (CEM) calculation described
in Ref.~\cite{Gavai:1994in}, NLO in the total cross section.  Thus the 
modifications
of the gluon distribution in the nucleus are similar in the CEM calculation
and the Glauber Monte Carlo using Eq.~(\ref{eqn:xq2}). This is demonstrated in 
Fig.~\ref{fig:kinematics}, where the EPS09 NLO gluon modifications obtained from 
Eq.~(\ref{eqn:xq2}) are compared with those obtained from the CEM calculation.
The results also agree with those found using PYTHIA~\cite{Nagle:2010ix}.

We fitted \sbr and the parameters derived from the \rt dependence of shadowing 
to the PHENIX data using a modified \chisqm function that properly accounts for 
all of the experimental uncertainties~\cite{Adare:2008cg}:
\begin{eqnarray}
\overline{\chi}^{2} = (\sum_{i=1}^k \frac{[{{\rdau}_i} + \epsilon_{{\rm B}_{i}} 
\sigma_{{\rm B}_i} + \epsilon_{\rm C} \sigma_{{\rm C}_i} - 
\rdau_i(M_{\rm shad},\sbr)]^2}{\overline{\sigma}_{{\rm A}_i}}) \nonumber \\
+ {\epsilon_{\rm B}}^2 + {\epsilon_{\rm s}}^2 + {\epsilon_{\rm C}}^2,
\label{eqn:chisq}
\end{eqnarray}		
\begin{equation}
\overline{\sigma}_{{\rm A}_i} = \sigma_{{\rm A}_i} \left(\frac{{{\rdau}_i} + 
\epsilon_{{\rm B}_i} \sigma_{{\rm B}_i} +\epsilon_{\rm C} 
\sigma_{{\rm C}_i}}{{{\rdau}_i}}\right),
\label{eqn:chisq_Ai}
\end{equation}
\begin{equation}
\epsilon_{{\rm B}_i} = \epsilon_{\rm B} + \epsilon_{\rm s}\left(1 - 2\frac{\langle 
\Lambda(\rt) \rangle_i - \langle \Lambda(\rt) \rangle_1}{\langle \Lambda(\rt) 
\rangle_{k} - \langle \Lambda(\rt) \rangle_1}\right),
\label{eqn:bi}
\end{equation}
where $i$ is the index of the centrality bin, $k$ is the number of centrality 
bins, ${{\rdau}_i}$ is the data point,
$\rdau_i({M_{\rm shad},\sbr})$ is the model calculation for the trial values of 
the absorption cross section and shadowing prescription,
$\sigma_{{\rm A}_i({\rm B}_i,{\rm C}_i)}$ are the type A (point to point), B 
(correlated systematic), and C (global) 
uncertainties on the data point. The effect of systematic uncertainties is 
included by moving the data points through $\pm 3\sigma$ in the type B and 
type C uncertainties, while taking an appropriate \chisqm penalty.  For each 
trial, $\epsilon_{{\rm B}({\rm C})}$ is the fraction of one standard deviation by 
which the data point moves. Note that Eq.~(\ref{eqn:bi}) contains a term that 
allows for some anti-correlation in the type B uncertainties. Here we allow 
the type B uncertainties to be linearly correlated
about the center of the distribution, {\it i.e.} $+\epsilon_s$ in the first
point and $-\epsilon_s$ in the last, and include a
corresponding penalty for this correlation. The value of $\epsilon_{s}$ was 
varied in the range $\pm 3\epsilon_{B}$. 
Although this is a reasonable prescription, the amount of correlation in
the type B uncertainty is unknown and could vary. However the type B
uncertainty is small ($\sim 2$\%) and therefore any 
correlation should have a negligible effect on the end result when
compared to the type A and C uncertainties.

Individual parameter uncertainties were evaluated by finding the
values at which \chisqm 
increased by 1.0 with all other parameters re-optimized.

To begin, we assume that the shadowing strength is
proportional to the nuclear thickness at impact parameter \rt raised
to a power $n$, $T_A^n(r_T)$,
\begin{eqnarray}
M_{\rm shad} = 1 - (1 - \rg(x,Q^2)\,)\bigg( \frac{T_A^n(r_T)}{a(n)} \bigg) \, \, .
\label{eqn:modshad_power}
\end{eqnarray}
Here $\rg(x,Q^2)$ is the EPS09 NLO gluon modification 
and the normalization factor, $a(n)$, is adjusted so that the integral over 
all impact parameters returns the average 
EPS09 modification. 
The power $n$ was allowed to be unphysically
large, $n \leq 50$, while the modification was constrained to be positive.
By allowing such arbitrarily large values of $n$, we can
test the sensitivity of the data to the centrality dependence of the
shadowing.  

The results of fitting with Eq.~(\ref{eqn:modshad_power}), described
in the next section, 
suggested that using a step function onset of shadowing,
including a radius, $R$, and a diffuseness, $d$, parameter,
\begin{eqnarray}
M_{\rm shad} = 1 - \bigg( \frac{1-\rg(x,Q^2)}{a(R,d)(1 + \exp((r_{_{T}} - R)/d))} \bigg) \,\, .
\label{eqn:modshad_sharp}
\end{eqnarray}
would be more appropriate. Again, the normalization factor  $a(R,d)$
is adjusted so that the integral over all impact parameters returns
the average EPS09 modification.  Thus,
Eq.~(\ref{eqn:modshad_sharp}) was the second form of the \rt
dependence of shadowing assumed in this work.

\section{Fitting Results}

\begin{figure}[htb]
 \includegraphics[width=1.0\linewidth]{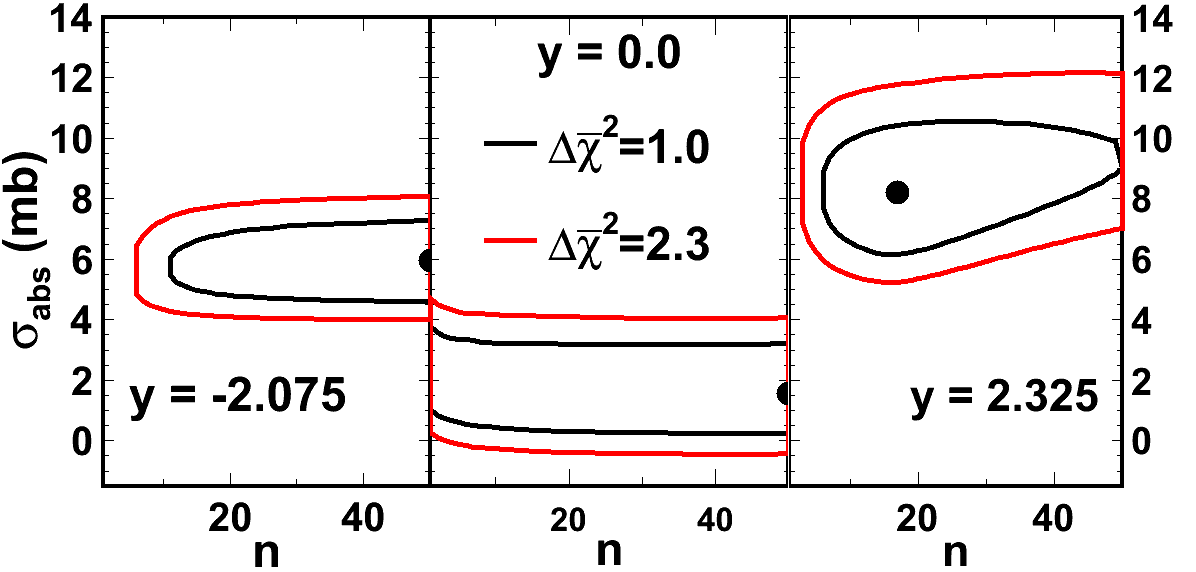}
 \caption{ The \chisqm distributions for the most backward, mid and most 
forward rapidities when \sbr and $n$ are optimized separately at each rapidity. 
   }
 \label{fig:three_panel_thick_contours}
\end{figure}

As a first step, the fits using Eq.~(\ref{eqn:modshad_power}) were made by determining 
the values of both \sbr and $n$ completely independently at each 
rapidity. This first step is not quite realistic because it ignores the correlations
among some of the systematic uncertainties within each of the three 
spectrometer arms. 
However it provides an indication of how the centrality dependence
varies with rapidity.
The \chisqm contours in \sbr and $n$ corresponding to $\Delta\chisqm =
1.0$ and 2.3 are shown  
in Fig.~\ref{fig:three_panel_thick_contours} for the most backward rapidity, 
midrapidity, and the most forward rapidity. At midrapidity, the fits 
are insensitive to $n$ because the shadowing effects are weak. At the
most forward and backward rapidities the optimum $n$ is large, $n \geq 10$, 
indicating that the data require a strongly nonlinear onset of shadowing or 
antishadowing as a function of impact parameter. Additionally, there is 
relatively little correlation, and thus little ambiguity, between $n$ and \sbr 
for $n$ greater than a few.

\begin{figure}[htb]
  \includegraphics[width=1.0\linewidth]{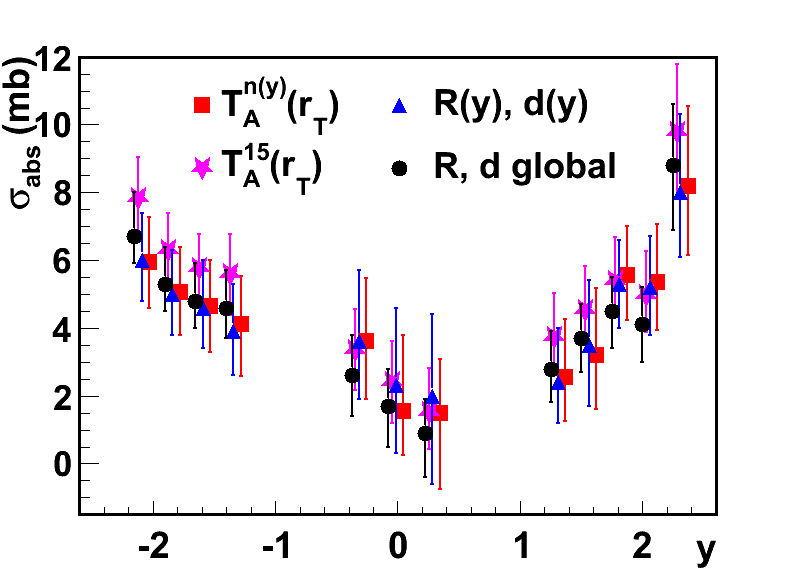}
 \caption{ The optimum values of and uncertainties on \sbr as a function of 
rapidity obtained from the four fits described in the text. 
For clarity, the rapidities  for each fit are slightly offset.
   }
 \label{fig:sigma_comparison}
\end{figure}

The optimum value of \sbr
and the corresponding uncertainty at each rapidity is shown by the red squares 
in Fig.~\ref{fig:sigma_comparison}. The \sbr values are reasonably well 
defined, with a minimum near midrapidity.
Because the data at all rapidities are consistent with $n\geq 5$, we
repeated the fit assuming a rapidity-independent value of $n$. For
this global fit, as well as for the one discussed later,
the systematic uncertainties correlated within each
experimental arm were handled correctly by being forced 
to move together. The optimum global power was $n=15^{+5}_{-4}$. The $\chisqm/{\rm dof}$ was 
1.94 overall. The best fit values of \sbr with $n=15$ are shown as stars in 
Fig.~\ref{fig:sigma_comparison}.

The fits favor (or, at midrapidity, are consistent with) a shadowing 
modification that is negligible at large \rt but turns on sharply as 
\rt decreases below $\sim 2- 3$~fm.  
This behavior suggested that the step function onset of shadowing described by
Eq.~(\ref{eqn:modshad_sharp}) would be more appropriate.  

For fits employing Eq.~(\ref{eqn:modshad_sharp}), we also initially
fit the parameters $R$ and $d$, along with \sbr, independently at each rapidity. 
The optimum \sbr values, the triangles in
Fig.~\ref{fig:sigma_comparison}, are very similar to those obtained from the 
earlier fits employing $T_A^n(r_T)$.  
The fits favor $R$ values of about half the Au radius, $R \leq 3.5$~fm.
While they also favor a small diffuseness parameter, they are
relatively insensitive to the value of $d$.

\begin{figure}[htb]
  \includegraphics[width=1.0\linewidth]{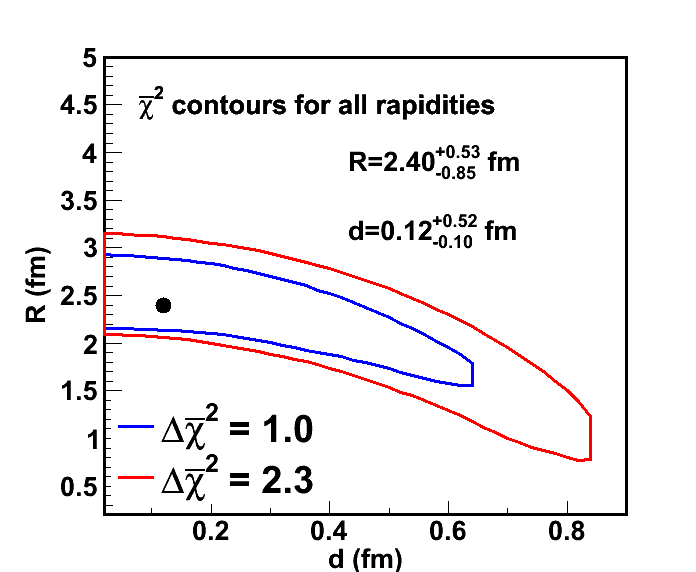}
 \caption{The \chisqm contours obtained from the fits with 
global values of $R$ and $d$, Eq.~(\protect\ref{eqn:modshad_sharp}). The
uncertainties in $R$ and $d$ are taken from the maximum extent of the 
$\Delta\chisqm = 1$ contour.}
 \label{fig:chisq_cont_allraps_R_a}
\end{figure}

\begin{figure}[htb]
  \includegraphics[width=1.0\linewidth]{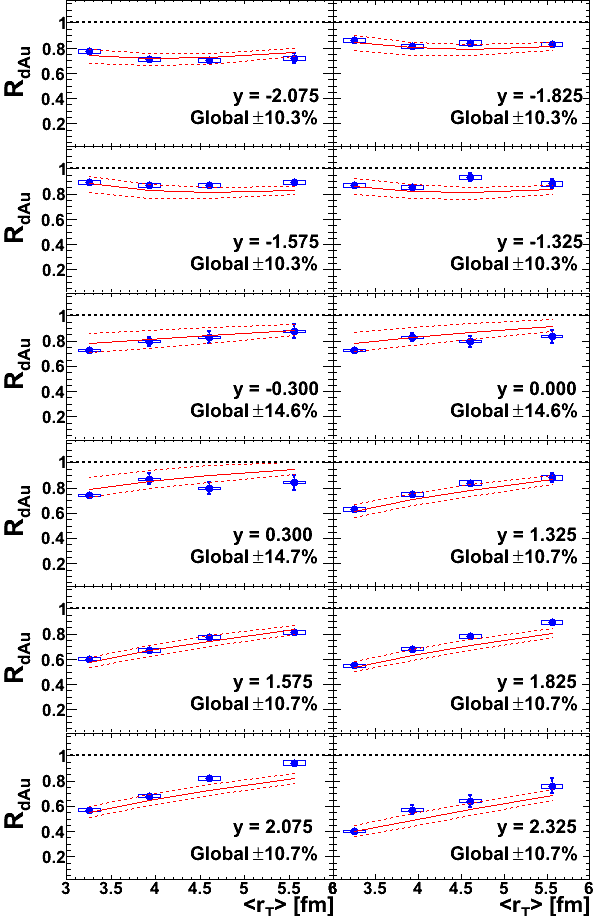}
 \caption{ Comparison to the data of the best fits with global $R$ and
   $d$ values. 
   }
 \label{fig:fitted_curves_step}
\end{figure}

Finally, we fit \sbr at each rapidity while requiring a global fit to
the values of $R$ and $d$.  The $\chisqm/{\rm dof}$ was 1.96 over all 
values of $y$.  The best fit \sbr at each rapidity 
are shown as circles in Fig.~\ref{fig:sigma_comparison}. 
The \chisqm contours in $R$ and $d$ are shown in 
Fig.~\ref{fig:chisq_cont_allraps_R_a}. 
The optimum global parameter values are 
$R=2.4^{+0.53}_{-0.85}$ fm and $d=0.12^{+0.52}_{-0.10}$ fm, where
the uncertainties are obtained from the maximum extent of the 
$\Delta\chisqm=1.0$ contour.
The fit results are compared with the measured \rdau as a function of \rt in 
Fig.~\ref{fig:fitted_curves_step} where the mean \rt values are the averages
obtained from the Glauber model for the four PHENIX centrality bins.
The dashed curves indicate the uncertainty in \rdau due
to the uncertainty in \sbr.
Because \chisqm includes the global uncertainties on the data, the best fit 
values may be slightly vertically offset in order to achieve the best
overall \chisqm.

\section{Discussion of Results}

This study indicates that, due to their very different \rt dependence,
there is little ambiguity between modifications due to shadowing and
the effective absorption cross section \sbr. For perspective, we first
compare the modifications due to shadowing 
and the effective absorption cross section calculated from
our fit parameters. Then we discuss the results for
effective absorption and shadowing, which presumably reflect different physics
processes, separately. 

\subsection{Relative contributions of shadowing and absorption}
\label{sec:Relative contributions to the modification from shadowing and absorption}

\begin{figure}[htb!]
 \includegraphics[width=1.0\linewidth]{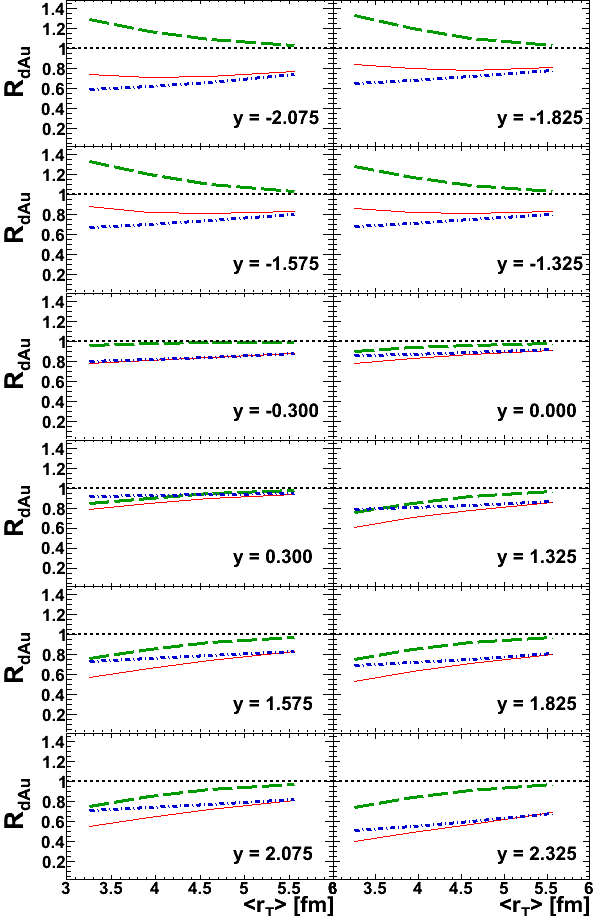}	
\caption{ The nuclear modifications averaged over the PHENIX
centrality resolution. The modification calculated from the fitted
\sbr alone is shown by the blue dot-dashed curve. The modification due
to shadowing, calculated with Eq.~(\ref{eqn:modshad_sharp}) using the
best fit global values of $R$ and $d$, is shown by the green dashed
curve. The product of the two effects, the overall calculated nuclear 
modification, is shown as the red solid curve.
}
 \label{fig:relative_mod_sigma_shadowing}
\end{figure}

It is of interest to contrast the modifications due to shadowing with those
due to the effective absorption cross section. This is done in 
Fig.~\ref{fig:relative_mod_sigma_shadowing},
after averaging over the PHENIX centrality resolution. 

We find that the nuclear modification due to the effective 
absorption cross section is typically larger than that due to shadowing
($R_{\rm dAu}$ is smaller), 
even at forward and backward rapidities, where shadowing effects are 
strongest.  However, the \rt dependence is stronger for shadowing and the
overall \rt dependence of $R_{\rm dAu}$ more closely follows that for
shadowing alone except at $y = -0.3$ where $R_{\rm dAu}\sim 1$ due to
shadowing.

\subsection{Effective absorption cross section}

The \sbr values obtained from fits employing different assumptions for
the centrality dependence of shadowing 
are all compared in Fig.~\ref{fig:sigma_comparison}. 
The fitted values of \sbr are well defined and are essentially independent of 
the \rt dependence assumed for the shadowing prescription, suggesting that the
effects of shadowing can be separated from those of 
physics processes that contribute to the value of the effective
absorption cross section extracted for the \ccbar pair. For specificity, in the 
following discussion  we use the \sbr values found using 
Eq.~(\ref{eqn:modshad_sharp}) with the values of $R$ and $d$ 
obtained from the global fit.

In practice, the effective absorption cross section encapsulates any 
physical process that reduces the \jpsi yield with an approximately 
exponential dependence on nuclear thickness. A mechanism that is 
linear in the nuclear thickness would be indistinguishable from
exponential in this study and would contribute to \sbr. Thus \sbr must 
contain a contribution from the reduction in \jpsi yield 
caused by the breakup of bound \ccbar pairs in collisions with Au nucleons that 
pass through the production point 
after the hard scattering. Additionally, it would likely contain the effects of 
energy loss, resulting in an effective rapidity shift \cite{Kharzeev:1993qd,Arleo:2012rs}.

\begin{figure}[htb]
  \includegraphics[width=1.0\linewidth]{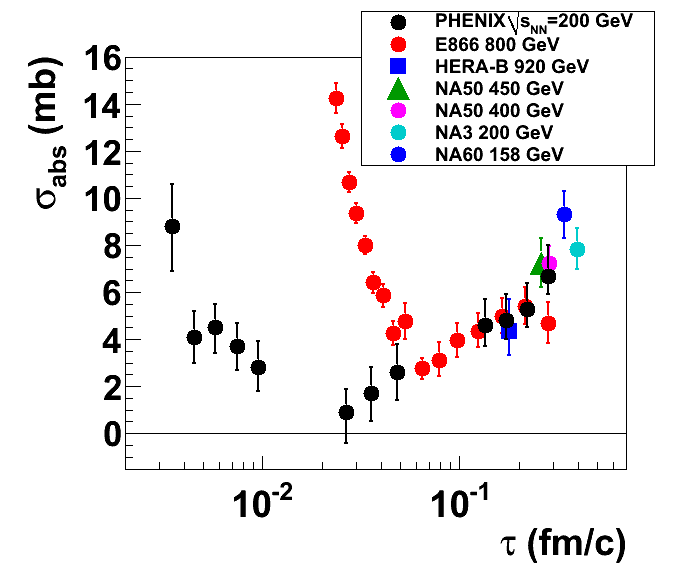}
 \caption{ 
The effective \ccbar absorption cross section \sbr as a function of the proper
time spent in the nucleus, $\tau$. The times are extracted fom
experiments over a range of energies.  
Those from the PHENIX $\sqrt{s_{_{NN}}} =200$ GeV d+Au data were
obtained after correcting for shadowing using the EPS09 NLO
parameterization (this work).  The values extracted from fixed target
experiments were corrected for shadowing using the EKS98
parameterization. The fixed target $p+A$ data used in  in 
Ref.~\protect\cite{Lourenco:2008sk} were from E866 at 800 
 GeV~\protect\cite{Leitch:1999ea}; HERA-B at 920 
GeV~\protect\cite{Abt:2008ya}; NA50 at 400 GeV
\protect\cite{Alessandro:2006jt} and 450 
GeV~\protect\cite{Alessandro:2003pc};  and NA3 at 200 
GeV~\protect\cite{Badier:1983dg}.  Those for NA60 at 158
 GeV were extracted in \protect\cite{Arnaldi:2010ky}.
}
 \label{fig:sigma-tau}
\end{figure}

A number of authors have pointed out that the \ccbar-nucleon cross section 
is expected to depend strongly on the 
size of the \ccbar pair as it expands to a fully formed meson~\cite{Farrar:1990ei,Blaizot:1989de,Gavin:1990gm,Arleo:1999af}. 
Therefore the proper time (in the frame of the \ccbar)  over
which the pair can collide with target nucleons should have an
effect on the apparent magnitude of the absorption cross section.  The
asymptotic cross section should be independent of time.
Figure~\ref{fig:sigma-tau} shows
the values of \sbr extracted from the PHENIX data as a function of
the proper time spent by the \ccbar in the target nucleus 
\begin{eqnarray}
\tau =  \frac{\beta_z L}{\gamma} \, \, ,
\label{eqn:tau}
\end{eqnarray}
where $L$ is half of the target thickness, averaged over all impact
parameters, and $\beta_z$ is the longitudinal 
velocity of the \ccbar relative to the target nucleus.
The Lorentz factor $\gamma$ for the \ccbar in the frame of the target 
nucleus converts the nuclear crossing time into the proper time for the \ccbar.
The measured mean transverse momentum, $\langle \pt \rangle$, was used when 
calculating $\gamma$ at each rapidity. The value of $L$
for a gold nucleus was taken from the Glauber model used in this work. The 
observed \jpsi yield includes feed-down from higher charmonium states. Following~\cite{Arleo:1999af}
we assume that the intermediate \ccbar and all charmonia states have an average mass of 
3.4 GeV/$c^2$ when calculating $\tau$.

The data display different behaviors as a function of $\tau$.  Thus,
in the next two subsections, we discuss the regions $\tau > 0.02$
fm/$c$ (higher $\tau$) and $\tau < 0.02$ fm/$c$ (lower $\tau$)
separately.

\subsubsection{Higher $\tau$ region}

It is notable that the mid and backward rapidity values of \sbr extracted from the PHENIX data increase as $\tau$ increases from 0.02 fm/$c$ to
0.3 fm/$c$, which is approximately the time scale over which a color
singlet $c \overline c$ 
expands to its final size~\cite{Arleo:1999af}. This is 
suggestive of the expected increase in \sbr with time spent in the
nucleus. On the other hand, 
the values of \sbr begin to increase again as $\tau$ drops below $\sim
0.01$ fm/$c$. Since the \ccbar formation time is 
expected to be $\sim 0.05$ fm/$c$~\cite{Arleo:1999af}, this increase
of \sbr with decreasing $\tau$ (increasing rapidity) is
presumably of a different physical origin.

If the behavior of \sbr at larger values of $\tau$ is dominated by the 
time the \ccbar spends in the nucleus, then the $\tau$
dependence should be independent of center-of-mass 
energy. 

Values of \sbr have been extracted from fixed target data, after 
correcting for shadowing, at six energies in the range $\sqsn=17.3 - 41.6$ 
GeV~\cite{Lourenco:2008sk, Arnaldi:2009ph, Arnaldi:2010ky}. 
To be able to compare those \sbr values to our results, it was necessary 
to estimate the average $\tau$ value in
all six cases. That in turn requires an estimate of the $\langle \pt \rangle$ values. 
For HERA-B data at 920 GeV 
beam energy there is a parameterization of the \pt distribution~\cite{Abt:2008ya}
that provides values of both $\langle p_T \rangle$ and $\langle p_T^2 \rangle$.
Additionally, that paper contains a systematic comparison of data from 
collisions at beam energies of 450 GeV, 800 GeV and 920 GeV showing 
that $\langle p_T^2 \rangle$ is linear with the square of the collision energy.
Using that fact, and assuming that the ratio  
$\langle \pt \rangle / \sqrt{\langle p_T^2 \rangle}$ is approximately 
constant from  $\sqsn=17.3 - 41.6$, we were able to estimate the 
values of $\langle \pt \rangle$, and thus $\tau$, at all six energies. 

We have added to Fig.~\ref{fig:sigma-tau} the values of \sbr 
extracted in~\cite{Lourenco:2008sk} from fixed target $p+A$ data 
from E866 at 800 GeV;  HERA-B at 920 GeV; NA50 at 450 and 
400 GeV; and NA3 at 200 GeV. The latter four values were extracted 
at $y$=0, while the E866 data cover a wide range of rapidities. 
In all cases the data were corrected 
for shadowing using the EKS98 parameterization.
The error bars shown include systematic uncertainties. We consider 
those cross sections to be directly comparable to the \sbr values 
extracted here using EPS09 shadowing, since the EKS98 and EPS09
central gluon modifications are very similar~\cite{Eskola:2009uj}. 
We have also included the \sbr value extracted 
from the NA60 data for $y \sim 0.3$ at 158 
GeV~\cite{Arnaldi:2009ph, Arnaldi:2010ky}, where shadowing 
was also corrected for using the EKS98
parameterization. 

\begin{table}[ht]
\caption{Kinematic characteristics used to determine the average time, $\tau$, 
the $J/\psi$ spends in the nucleus for several experiments and targets, 
shown in Fig.~\ref{fig:sigma-tau}.}
\centering
\begin{tabular}{| c | c | c | c | c | c | c |c|}
\hline\hline
Experiment & $\sqrt{s_{_{NN}}} $ & $A$ & $y_{\rm beam}$ & $y_{\rm cm}$
& $L$ & $\langle \pt \rangle$ & $\tau$   
\\ [0.5ex] 
  & (GeV) &  &  & & (fm) &  GeV/$c$ & (fm/$c$)  \\ [0.5ex] 
\hline
PHENIX & 200  & Au  & 5.36  &  -2.08-2.32 &  4.36  & 1.90 & 0.283 - 0.0035  \\ 
HERA-B & 41.6 & W & 7.58 & 0.0 & 4.26 &  1.36 & 0.178 \\
E866 & 38.8 & W & 7.44 & -0.39-2.1  & 4.26 & 1.32 & 0.283 - 0.024  \\
NA50 & 29.1 &  W & 6.87 & 0.0 & 4.26 &  1.22 & 0.258  \\
NA50 & 27.4 & Pb  & 6.75 & 0.0 & 4.44 & 1.20 & 0.286  \\
NA3  & 19.4 & Pt & 6.06 & 0.0 & 4.34 & 1.14 & 0.396  \\
NA60 & 17.3 & Pb & 5.82 & 0.3 & 4.44 &  1.12 & 0.339  \\
\hline
\end{tabular}
\label{table:lvalues}
\end{table}

The average $L$, $\langle \pt \rangle$ and the kinematics characteristic of the
experiments used to calculate the $\tau$ values from
Eq.~(\ref{eqn:tau}), displayed in 
Fig.~\ref{fig:sigma-tau}, are given in Table~\ref{table:lvalues}.  
The W/Be ratios were used for
E866; W/C ratios for HERA-B; and Pt/H ratios for NA3.  
A range of targets was employed by the NA50 and NA60 Collaborations,  
and the value of $L$ for the heaviest target is shown in
Table~\ref{table:lvalues}. 
The $L$ values for all targets other than Au were obtained from the Au value 
assuming $A^{1/3}$ scaling.

For $\tau$ greater than 0.05 fm/$c$ the lower energy data 
seem to follow the same trend as those extracted from
PHENIX data. The data in Fig.~\ref{fig:sigma-tau} cover the energy range $17.3 \leq
\sqrt{s_{_{NN}}} \leq 200$ GeV. The
common behavior of \sbr with $\tau$ for $\tau > 0.05$ fm/$c$ across such a large
$\sqrt{s_{_{NN}}}$ range is striking. 

The results in Fig.~\ref{fig:sigma-tau} imply that, for $\tau > 0.05$ fm/$c$,
\sbr depends on the time the $c \overline c$ spends in the nucleus. 
Thus \sbr is a function of both the nuclear target mass and
impact parameter. This was not taken into account when 
extracting the values of \sbr included in Fig.~\ref{fig:sigma-tau}. 
However, we note that changing the average $L$ by a factor of two produces
only about a 1 mb change in \sbr.  Therefore averaging \sbr over 
a range of impact parameters and, in the case of the NA50 400 and 450
GeV data, over a range of targets, may be acceptable. 

A description of charmonium suppression by  nucleon absorption in
$p+A$ collisions proposed by Arleo {\it et al.}~\cite{Arleo:1999af}
may be illustrative in this large $\tau$ region.  In this approach,
the \ccbar pair, assumed to be initially formed in a color octet
state, neutralizes its color by gluon emission and expands to the
physical size of the meson. In cases where $\tau$ is short, the \ccbar 
travels through the target as a colored object.  When $\tau$ is long,
it travels through the target as an expanding or fully-formed color
singlet. In the latter case, the absorption cross section depends on $\tau$ 
due to the dependence of the $c\overline{c}$ radius on $\tau$. 

\begin{figure}[htb]
  \includegraphics[width=1.0\linewidth]{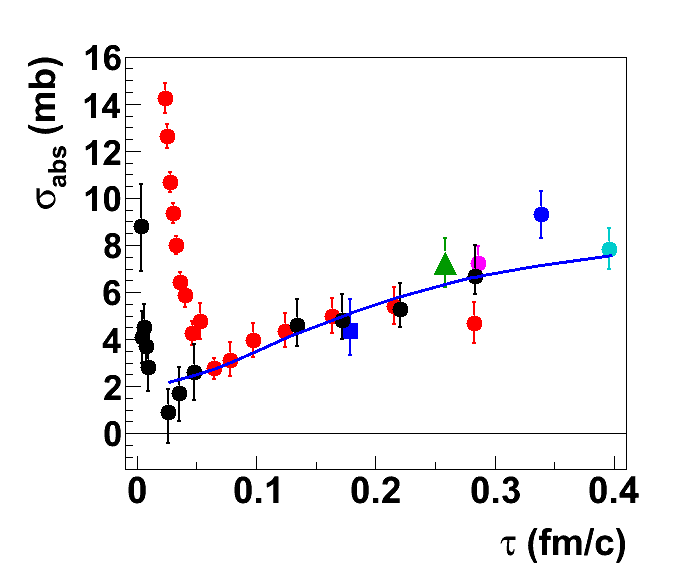}
 \caption{ 
 The same data as in Fig.~\ref{fig:sigma-tau} plotted on a linear
 scale compared to a calculation of the cross section based 
 on the model of Ref.~\protect\cite{Arleo:1999af}.
}
 \label{fig:sigma-tau-fit}
\end{figure}

When the \ccbar is still expanding, the $\tau$
dependence of \sbr was assumed to be~\cite{Arleo:1999af} 
\begin{eqnarray}
\sbr(\tau) = \sigma_{1} \bigg( \frac{\sqrt{s}}{10~\rm{GeV}}\bigg)^{0.4} \bigg(\frac{r_{c\overline{c}}(\tau)}{r_{\rm{J}/\psi}}\bigg)^{2} ,
\label{eqn:sbr-tau-fit}
\end{eqnarray}
where $\sigma_{1}$ is the cross section for destruction of a fully
formed \jpsi in an interaction with a nucleon at $\sqs = 10$ GeV. The
time dependence of the \ccbar radius was taken to be
\begin{eqnarray}
r(\tau) = r_{0}+v_{c\overline{c}} \tau \,\,\, ({\rm if}~r(\tau) <
r_{\psi})\, \, .
\label{eqn:r-tau1}
\end{eqnarray}


In Ref.~\cite{Arleo:1999af} it was assumed that the observed \jpsi yield was a
combination of direct \jpsi production and feed down from 
the $\psi^\prime$ and $\chi_{c}$. The model parameters were extracted
from fits to the $A$ dependence of the E866 \jpsi data. However, no
correction was made for shadowing.  Thus the parameters obtained in 
Ref.~\cite{Arleo:1999af} are not appropriate for the
shadowing-corrected \sbr values we extract from the PHENIX data, or 
the shadowing corrected \sbr values extracted from the lower energy data. 
Therefore we have fitted the parameters in Eqs.~(\ref{eqn:sbr-tau-fit})
and (\ref{eqn:r-tau1}) to the subset of the cross section data shown in
Fig.~\ref{fig:sigma-tau} with $\tau > 0.05$ fm/$c$ ($\tau > 0.02$ fm/$c$ for PHENIX).
At each energy and rapidity, the calculated \sbr was averaged 
over \rt using the distribution of nucleon-nucleon collisions obtained from the Glauber
simulation, and averaged over the (longitudinal) $z$ dimension using the same 
Woods Saxon distribution as was used in the Glauber simulation.

Because the data being fitted 
are from a variety of targets and a wide range of collision energies, we have 
made some simplifying approximations. First, we calculated the $\tau$ values only for 
the $\langle p_T \rangle$ at each energy (see Table~\ref{table:lvalues}), rather than 
averaged over the \pt distribution. This assumption was checked
for the PHENIX 200 GeV case by averaging over the full measured \pt distribution in the Glauber model, 
and it was found to give the same result for the average $\tau$ and average 
\sbr to much better than 1\%. Second, we assumed that the distribution
of nucleon-nucleon collisions for all targets (mass range 184 to 208) was adequately described by that 
for Au (mass 197). For the lower energy data, the average $\tau$ values obtained 
using this assumption differed by less than 2.5\% from those obtained using the $A^{1/3}$ scaled 
average length given in Table~\ref{table:lvalues}.

The fraction of \jpsi's arising from direct production together with $\psi^\prime$ 
and $\chi_c$ feed down were taken to be 58\%, 10\% and 32\% respectively, see
Ref.~\cite{Adare:2011vq}. The radii of the three \ccbar states were
assumed to be 0.43 fm for the \jpsi, 0.87 fm for the $\psi^\prime$ and
0.67 fm for the $\chi_{c}$, as in Ref.~\cite{Arleo:1999af}. 

The best fit of Eqs.~(\ref{eqn:sbr-tau-fit}) and (\ref{eqn:r-tau1}) to
the high $\tau$ data is shown in Fig.~\ref{fig:sigma-tau-fit}. The optimum
values of the parameters are  $\sigma_{1} = 7.2$ mb,  $r_{0} = 0.16$ fm
and $v_{c\overline{c}}=1.0$. The curve resulting from the fit describes 
all of the high $\tau$ \sbr values very well, with a $\chi^{2}/{\rm dof}$ of 0.94.
We note that the fitted 
value of $\sigma_{1}$ is considerably larger than that obtained in 
Ref.~\cite{Arleo:1999af}. The difference arises, at least in part, due to the
substantial antishadowing correction from the EPS09 and EKS98 
parameterizations used to extract the \sbr values from data at higher $\tau$,
which results in larger \sbr than those extracted from fits without any 
shadowing correction \cite{Lourenco:2008sk}. 

\begin{figure}[htb]
  \includegraphics[width=1.0\linewidth]{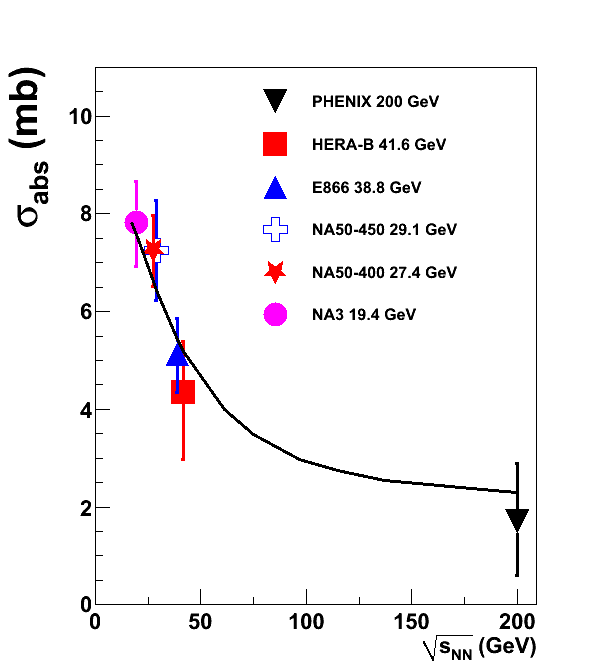}
 \caption{ 
Measured \sbr values at $y=0$ compared with the energy dependence calculated from
the best fit to \sbr versus $\tau$ shown in Fig.~\ref{fig:sigma-tau-fit}. The 200 GeV data point
is from the present work, all others are from~\cite{Lourenco:2008sk}.
}
 \label{fig:sigma_vs_roots_midrapidity}
\end{figure}

The collision energy dependence of \sbr at $y=0$ obtained from the 
fit shown in Fig.~\ref{fig:sigma-tau-fit} is compared with data in 
Fig.~\ref{fig:sigma_vs_roots_midrapidity}. 

We conclude that a model of time-dependent nucleon absorption
such as that of Ref.~\cite{Arleo:1999af} is capable of describing the 
trend of the \sbr values extracted from the PHENIX data at mid- and 
backward rapidity, as well as those extracted from lower energy data
for $\tau > 0.05$ fm/$c$.
 
\subsubsection{Lower $\tau$ region}

The other striking feature of the $\tau$ dependence of \sbr is the
departure from $\tau$ scaling when the \ccbar spends only a short
time traversing the nucleus. 
The results extracted from the PHENIX and E866 data both
show a strong increase in \sbr at small $\tau$, starting at $\tau \sim
0.05$ fm/$c$ in the E866 case and at
$\tau \sim 0.02$ fm/$c$ for PHENIX. These times are smaller than or 
comparable to the \ccbar formation time and the color neutralization
time.  Several models have been proposed which might explain 
increased \jpsi suppression at forward rapidity. 

Arleo {\it et al.}~\cite{Arleo:1999af} assumed that the increased E866
cross section at forward $x_F$ (rapidity) was due to the interaction
of nucleons with the color octet state. Here the $\tau$-independent 
cross section is $22.3 (\sqrt{s_{NN}}/10)^{0.4}$ mb for $\tau < 0.02$ fm$/c$, the 
color neutralization time, a parameter of their model. Such behavior
leads to increasing suppression with decreasing $\tau$ because a
smaller fraction of \ccbar pairs escape before the color neutralization
time is reached.  In this picture, the increased suppression at forward rapidity is
due to breakup by nucleons.  However, in this approach, \sbr should
also scale with $\tau$ which clearly does not hold for the E866 and
PHENIX data at small $\tau$, as seen in Fig.~\ref{fig:sigma-tau}.
Therefore, other mechanisms need to be considered.

In a recent paper, Arleo and Peigne~\cite{Arleo:2012hn,Arleo:2012rs} describe
forward rapidity data from NA3, E866 and PHENIX using a model 
of parton energy loss in cold nuclear matter. The \ccbar pair is assumed to be in an essentially point-like
long-lived color octet state, $\tau_{\rm octet} \gg \tau_{c \overline c}$, where $\tau_{c \overline c} \sim 1/M$, the
inverse of the pair mass, with a lifetime less than that of the
formation time of the 
$J/\psi$.  In this picture, gluon splitting, $g \rightarrow Q \overline
Q$, is followed by scattering with a gluon in the target nucleus.
The scattering with the target gluon is similar to Bethe-Heitler radiation off
a fast color octet undergoing transverse momentum broadening by scattering
with a coherent color field.  Since the scattering is from a target gluon,
Drell-Yan production, quarkonium photoproduction and nDIS are unaffected by 
this radiation.
They find $\Delta E \propto E$, as in Refs.~\cite{Gavin:1991qk,Johnson:2001xfa},
in contradiction to the energy independent bound derived in 
Ref.~\cite{Brodsky:1992nq} which neglected nuclear broadening in the 
final state.  Arleo and Peigne suggest that the bound holds only for abelian
models and not in the non-abelian case of QCD.
They fit the energy loss parameter $\hat{q}$ to the E866 $J/\psi$ data as
a function of $x_F$ and use that same parameter to calculate results at 
different energies.  They assume $2 \rightarrow 1$ kinematics and fit
the $pp$ production cross section to a power law in $x$, 
$d\sigma_{pp}/dx_F \propto (1-x)^n/x$ rather than making any
assumptions about
the quarkonium production mechanism.  Since the parameter governing
energy loss is related to the transverse momentum broadening of the state,
$l_T^2 \simeq \hat{q} L$, there is some centrality dependence that can
be introduced into the model in future work.

Arleo and Peigne added suppression of the \jpsi yield by either gluon 
saturation or standard shadowing parameterizations, including
EPS09. They did not, however, incorporate nuclear absorption, which
they noted may become important at backward rapidity.
Adding EPS09 shadowing led to very good agreement with the PHENIX \rdau data 
at forward rapidity and midrapidity.  However, it  resulted in a
considerable overprediction of the 
data (underpredicting the suppression) at backward rapidity.
Their backward rapidity result seems to be consistent with our present
results, where we find evidence that at backward
rapidity the modification is well described by a large absorption
cross section together with antishadowing.

Thus in the small $\tau$ (forward rapidity) region, energy loss 
effects may explain the rise in the effective absorption cross section 
observed here. If that is the case, the large effective absorption cross 
section we obtain at small $\tau$ is not due to \ccbar breakup, but 
is rather an energy loss-induced shift in the rapidity of the 
detected \jpsi. The effect is expected to depend on the square root of the 
thickness~\cite{Arleo:2012hn}, rather than the exponential dependence
implied by our use of an effective absorption cross section. 
However the $\sqrt{T_{A}}$ dependence is sufficiently close to exponential 
that our fitting procedure is unable to discriminate between them.

\subsection{Shadowing}
\label{sec:shadowing}

\begin{figure}[htb!]
 \includegraphics[width=1.0\linewidth]{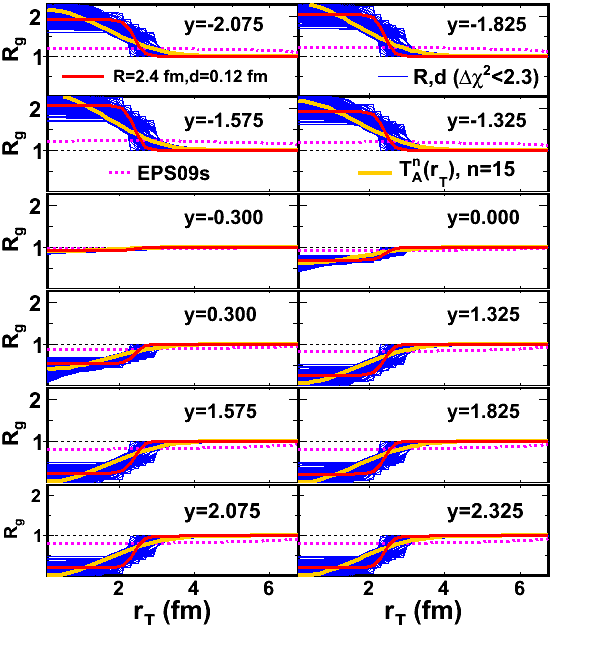}	
\caption{The gluon modification from the best fit 
global $R$ and $d$ parameters (solid red line), along with the 
modifications from all combinations of $R$ and $d$ that fall within the 
$\Delta\chisqm=2.3$ contour in Fig.~\protect\ref{fig:chisq_cont_allraps_R_a} 
(thin blue lines).  The modification from the best fit global analysis of
$T_A^n(r_T)$ ($n=15$) is shown by the solid orange line. 
The  dashed magenta line is the recently released EPS09s NLO impact
parameter dependence \protect\cite{Helenius:2012wd}. 
}
 \label{fig:mod_allraps_step_fit}
\end{figure}

The gluon modification obtained from the EPS09 NLO 
parameterization using Eq.~(\ref{eqn:modshad_sharp}), 
with global fit values 
of $R$ and $d$ as a function of \rt, is shown by the solid red line  in
Fig.~\ref{fig:mod_allraps_step_fit}.
The effect of the combined uncertainty in
$R$ and $d$ can be visualized by plotting the modifications for all
combinations of $R$ and $d$ that produce a \chisqm value inside the 
$\Delta\chisqm = 2.3$ contour~\cite{numericalrecipes} 
(see Fig.~\ref{fig:chisq_cont_allraps_R_a}). 
These are represented by the
thin blue lines in Fig.~\ref{fig:mod_allraps_step_fit}.
In all cases, the calculated modification is significant only for $\rt 
\lesssim 3$~fm. Therefore, we conclude that the data constrain
the nuclear modification to be important only at small \rt. 
The modification obtained with the best fit global power, $T_A^{n}(r_T)$, $n=15$,
is shown as the solid orange line in Fig.~\ref{fig:mod_allraps_step_fit}.
Although there is some difference in the details at small \rt, albeit within
the uncertainties, the two
prescriptions give essentially the same values of $\chisqm/{\rm dof}$. 
Thus the data appear to be insensitive to the detailed shape of the 
modification at low \rt. This is because the d+Au centrality bins are 
wide and significantly overlap.  If the centrality bins were narrower, the sensitivity
to the centrality dependence could be increased.

We compare our results obtained from the fits to the $J/\psi$
data with those given by the newly-available impact parameter dependent EPS09s (NLO)
set~\cite{Helenius:2012wd}, shown by the dotted magenta line in 
Fig.~\ref{fig:mod_allraps_step_fit}. The EPS09s result has a much weaker
dependence on \rt than obtained from our fits. A study of the target mass 
systematics of \jpsi production in \dA collisions at RHIC may shed light
on the source of this pronounced difference.

\subsection{Effect of nonlinear shadowing on \sbr}

In Fig.~\ref{fig:sigma-tau}  we compare the \sbr values  
extracted from the PHENIX data in this work with the \sbr values 
extracted from centrality-integrated lower energy 
data~\cite{Lourenco:2008sk} or, for NA60 $p$+Pb, from fits to 
centrality-dependent data~\cite{Arnaldi:2010ky} 
that assumed a linear dependence of shadowing on nuclear thickness. 
Thus when comparing our values of \sbr with those obtained at 
lower energy, it is important to understand if the
\sbr values extracted from PHENIX data here depend strongly on the centrality 
dependence assumed for the shadowing.

\begin{figure}[htb!]
 \includegraphics[width=1.0\linewidth]{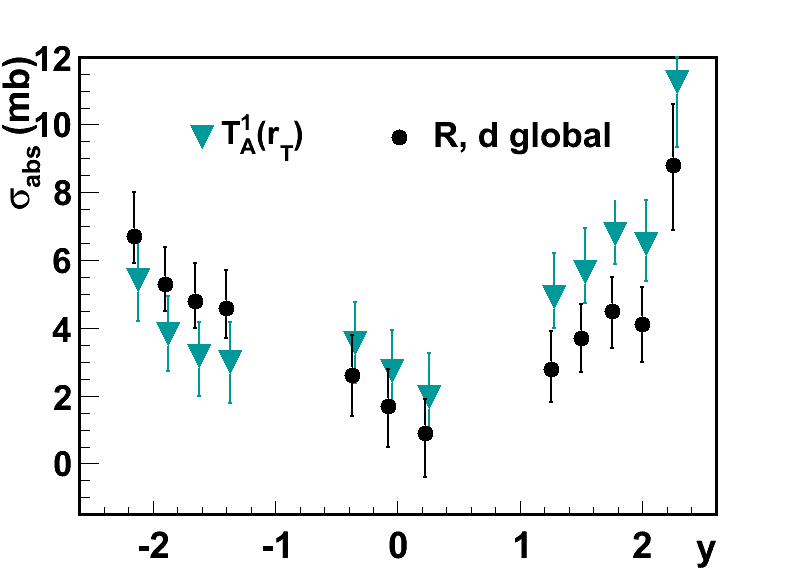}	
\caption{Comparison of the \sbr values extracted from the PHENIX data 
assuming a linear thickness dependence for the shadowing with those 
extracted using Eq.~(\ref{eqn:modshad_sharp}) assuming global values of 
$R$ and $d$.
}
 \label{fig:compare_Rd_with_n1}
\end{figure}

Fig.~\ref{fig:compare_Rd_with_n1} compares the \sbr values extracted 
from the PHENIX data using 
Eq.~(\ref{eqn:modshad_sharp}) with global values of $R$ and $d$, 
the values used in Fig.~\ref{fig:sigma-tau}, with \sbr values extracted 
assuming a linear dependence of shadowing on nuclear thickness. While 
there are some differences in the extracted \sbr if the shadowing has a
linear thickness dependence, they are not large enough to affect the 
conclusions drawn from Fig.~\ref{fig:sigma-tau}.

We emphasize that the shadowing description of the PHENIX data at both
backward and forward rapidity is much poorer when a linear thickness 
dependence is assumed. At the four backward rapidities, the \chisqm worsens 
by 4.3, while at the five forward rapidities it worsens by 47.6. At 
midrapidity, where the shadowing is weak, the \chisqm worsens by only 0.8.

\section{Summary and Conclusions}

In summary, we have fitted the centrality and rapidity dependent PHENIX 
$\sqsn=200$ GeV d+Au \jpsi data with Glauber calculations employing an
effective absorption cross section, \sbr, with several prescriptions for the
impact parameter dependence of the EPS09 NLO central gluon shadowing 
parameterization. The fits properly account for all of the experimental 
systematic uncertainties. We find little ambiguity between 
\sbr and the functional form of the centrality dependence of shadowing. 

The values of \sbr exhibit a characteristic rapidity dependence, with a 
minimum at midrapidity. When plotted as a function of
the average time the $c\overline{c}$ spends in the nucleus, $\tau$, 
the \sbr values extracted 
from the PHENIX data at $\sqsn=200$ GeV and from lower energy data
with $17.3< \sqsn < 41.6$ GeV,  display a common $\tau$ dependence 
for $\tau > 0.05$ fm/$c$. In that $\tau$ range the cross section is very well
described when the data are fitted with a model in which
the $c\overline{c}$-nucleon cross section depends on
the size of a color neutral $c\overline{c}$ as it expands into a 
fully-formed meson.  Such a model naturally leads to scaling of 
\sbr with $\tau$. The best fit parameters provide 
an excellent description of the collision energy dependence of 
\sbr at $y=0$ from $\sqsn=20-200$ GeV.

As $\tau$ decreases below $\sim 0.02$ fm/$c$, the \sbr values extracted from 
the PHENIX data rise sharply. This $\tau$ range is smaller than 
the expected $c\overline{c}$ formation time and color neutralization
time, and reflects different physical processes than those active at
higher $\tau$. The \sbr values extracted from E866 data at
$\sqsn=41.6$ GeV also exhibit a 
sharp rise beginning at $\tau \sim 0.05$ fm/$c$. The PHENIX
and E866 data show no scaling with $\tau$ in the range $\tau < 0.05$ fm/$c$.
The present results at low $\tau$ (forward rapidity) seem to be 
consistent with energy loss of a color octet \ccbar state in cold 
nuclear matter~\cite{Arleo:2012hn,Arleo:2012rs}. If so, the fitted effective \sbr
values do not reflect breakup of the \ccbar pairs, but instead an energy-loss
induced rapidity shift of the \jpsi. 

The centrality dependence of shadowing extracted from the data turns on 
sharply for $r_T \leq 3$~fm, in significant
disagreement with the weaker $r_T$ dependence of EPS09s NLO.  Indeed, the EPS09s
dependence is somewhat weaker than the linear dependence on the thickness
function assumed in Ref.~\cite{Klein:2003dj}. A study of the target mass 
systematics of \jpsi production in \dA collisions at RHIC may shed light
on the source of this pronounced difference.

 While we have employed only
the central EPS09 set in our calculations, using all 31 EPS09 sets would not
affect our overall conclusions regarding the sharp turn on of shadowing with
$r_T$, only increase the uncertainty in the value of \sbr as a function of
rapidity. The use of alternative nPDF sets [2-5] would also
change $\sigma_{\rm abs}(y)$ without affecting the $r_T$ dependence of
shadowing.
The strong impact parameter dependence suggested here seems
to be in accord with the `hot spots' conjectured in a saturated medium of
high gluon density.  Such behavior at backward rapidity, in the antishadowing
region, is, however, at odds with the saturation picture and may more simply
suggest that shadowing effects are concentrated in the core of the nucleus
instead of throughout the nuclear volume.


%

\begin{acknowledgments}
The work of R. V.  was performed under the auspices of the U.S.\
Department of Energy by Lawrence Livermore National Laboratory under
Contract DE-AC52-07NA27344, and  supported in part by the JET collaboration.
The work of A.D.F and D.C.M. was supported in part by the National Science Foundation grant
number PHY-10-64819.

\end{acknowledgments}

\bibliography{CNM_dAu_paper}
\end{document}